\documentclass[letterpaper,times]{IONconf}
\usepackage{amsmath}

\usepackage{amssymb}
\usepackage{enumitem}

\usepackage[noend]{algpseudocode}
\usepackage{algorithm}

\usepackage{siunitx} % for significant figures and number formatting
\usepackage{booktabs} % for better looking tables

\usepackage{graphicx}
\usepackage[justification=centering]{caption}
\usepackage{subcaption}
\graphicspath{{figures/}}
\usepackage{url}
\usepackage{natbib}
\usepackage[hidelinks]{hyperref}
\newcommand{\reals}{{\mbox{\bf R}}}

\newcommand{\BEQ}{\begin{equation}}
\newcommand{\EEQ}{\end{equation}}
\newcommand{\BEAS}{\begin{eqnarray*}}
\newcommand{\EEAS}{\end{eqnarray*}}
\newcommand{\BEA}{\begin{eqnarray}}
\newcommand{\EEA}{\end{eqnarray}}    

\newcommand{\BBM}{\left[\begin{matrix}}
\newcommand{\EBM}{\end{matrix}\right]}

\newcommand{\BIT}{\begin{itemize}}
\newcommand{\EIT}{\end{itemize}}
\newcommand{\BNUM}{\begin{enumerate}}
\newcommand{\ENUM}{\end{enumerate}}

\newcommand{\argmin}{\mathop{\rm argmin}}

\newcommand{\eg}{{\it e.g.}}
\newcommand{\ie}{{\it i.e.}}

\newcommand{\PAR}[1]{\paragraph{#1.}}

\usepackage{fancyhdr}
\title{Large-Scale GNSS Spreading Code Optimization}

\author{
Alan Yang\textsuperscript{1}, 
Tara Mina\textsuperscript{2}, 
Stephen Boyd\textsuperscript{1}, 
Grace Gao\textsuperscript{1}
\vspace{1mm} \\%
\textsuperscript{1}\textit{Stanford University},
\textsuperscript{2}\textit{Georgia Institute of Technology}%
}

\begin{document}

\maketitle

% biography section. The * indicates a section excluded from numbering.
\section*{biography}

% Biographies are defined as follows:
% \biography{Author name}{author biography text}

\biography{Alan Yang}{is a Ph.D. candidate in the Department of Electrical
Engineering at Stanford University. He received his M.S. degree in Electrical
and Computer Engineering in 2021 from the University of Illinois at
Urbana-Champaign, where he also received his B.S. in 2018. His research
interests include convex optimization, control, and applications.}

\biography{Tara Mina}{is a research engineer at Georgia Tech in the Space
Exploration Analysis Laboratory. She received her Ph.D. in the Electrical
Engineering Department from Stanford University in 2023. Prior to that, she
obtained her B.S. in Electrical Engineering from Iowa State University in 2017
and her M.S. in Electrical and Computer Engineering from the University of
Illinois at Urbana-Champaign in 2019. Her research interests include secure
positioning, navigation, and timing (PNT), GNSS signal design, lunar PNT, and
optical navigation for space exploration.}

\biography{Stephen Boyd}{is the Samsung Professor of Engineering
and Professor of Electrical Engineering at Stanford University.
%His current interests include convex optimization applications in 
%control, machine learning, signal processing, and finance.
He received an AB degree in Mathematics, summa cum laude,
from Harvard University in 1980, and a PhD in EECS from
U. C. Berkeley in 1985.  
%He holds honorary doctorates from Royal Institute of Technology (KTH)
%and Catholic University of Louvain (UCL).
He is a fellow of IEEE, SIAM, INFORMS, and IFAC, a member of the US National
Academy of Engineering, a foreign member of the 
Chinese Academy of Engineering, and a foreign member of the National
Academy of Engineering of Korea.}

\biography{Grace Gao}{is an associate professor in the Department of Aeronautics
and Astronautics at Stanford University. Before joining Stanford University, she
was an assistant professor at University of Illinois at Urbana-Champaign. She
obtained her Ph.D. degree at Stanford University. Her research is on robust and
secure positioning, navigation, and timing with applications to manned and
unmanned aerial vehicles, autonomous driving cars, as well as space robotics.}

% The Abstract. The * indicates a section excluded from numbering.
\section*{Abstract}
We propose a bit-flip descent method for optimizing
binary spreading codes with large family sizes and long lengths, addressing the
challenges of large-scale code design in GNSS and emerging PNT applications. The
method iteratively flips code bits to improve the codes' auto- and
cross-correlation properties. In our proposed method, bits are selected by
sampling a small set of candidate bits and choosing the one that offers the best
improvement in performance. The method leverages the fact that incremental
impact of a bit flip on the auto- and cross-correlation may be efficiently
computed without recalculating the entire function. We apply this method to two
code design problems modeled after the GPS L1 C/A and Galileo E1 codes,
demonstrating rapid convergence to low-correlation codes. The proposed approach
offers a powerful tool for developing spreading codes that meet the demanding
requirements of modern and future satellite navigation systems.

% The introduction. Section numbering starts here.
\section{Introduction}

A global navigation satellite system (GNSS) comprises a constellation of
satellites that transmit signals to Earth. Each satellite's signal contains a
\emph{spreading code}, which is combined with navigation data and modulated onto
a carrier wave at a specific frequency for transmission. Because all signals are
transmitted at the same frequency, the spreading codes are designed to have low
autocorrelation and low cross-correlation. This enables the receiver to
distinguish signals from individual satellites and accurately determine their
timing \citep{MisraE2012}.

Traditionally, pseudorandom binary sequences generated using linear shift
registers have been employed as spreading codes. For instance, the Global
Positioning System (GPS) utilizes Gold codes \citep{Gold1967} and Weil codes
\citep{Legendre1808, Rushanan2007}. While these codes exhibit favorable
autocorrelation and cross-correlation properties, they are only available in
specific code lengths. Additionally, they are suboptimal because they cannot be
tailored to a given code family size, \ie, the number of codes required for a
particular application.

There has been recent significant interest in developing new spreading codes
\citep{WallnerAHR2007,winkel2011spreading,YangMG2023ion}, as well as codes for
wireless communications \citep{JiangKAKOV2020} and radar \citep{AlaeeMN2019}
using optimization techniques. These methods enable the design of spreading
codes with good autocorrelation and cross-correlation properties, tailored to
specific applications and not restricted by predefined code lengths. For
example, the European Union's Galileo constellation employs spreading codes
designed using a genetic algorithm \citep{WallnerAHR2007,WallnerAWHI2008}.
Additionally, the upcoming Navigation Technology Satellite-3 (NTS-3) will serve
as a platform for testing future GPS technologies, marking the first time a GPS
satellite signal can be fully reprogrammed while in orbit
\citep{ChapmanHA2020,NTS3}. Moreover, emerging Positioning, Navigation, and
Timing (PNT) applications, such as those in low Earth orbit (LEO) and Lunar PNT,
stand to benefit from custom-designed spreading codes \citep{Soualle2005,
ReidWELCGOW2020, israel2023lunanet}.

Spreading code design is a challenging combinatorial optimization problem.
Although numerous heuristic methods have been proposed to identify codes with
favorable correlation properties, few are well-suited for the large-scale code
design challenges encountered in GNSS applications
\citep{RodriguezWH2006,AlaeeMN2019,MinaG2022,YangMG2023}. The number of required
codes can be substantial, often in the hundreds, particularly for LEO PNT
applications. Additionally, code lengths can be considerable, sometimes reaching
tens of thousands, since longer codes generally offer better performance in low
signal-to-noise ratio environments.

In this work, we propose a bit-flip descent method suitable for optimizing
spreading codes with large code family sizes and long code lengths. In each
iteration of the method, a single bit, or code entry, is selected and flipped if
it improves the objective function. To ensure rapid convergence, the method
samples a small set of candidate bits and selects the one that offers the
greatest improvement in the objective. This approach is particularly effective
for large-scale code design problems, since the impact of flipping a single bit
on the objective function can be efficiently calculated without recomputing the
entire objective function \citep{winkel2011spreading}. Moreover, bit-flip
descent methods have been shown to perform as well as, if not better than, more
sophisticated methods for code design problems \citep{YangMG2023ion}. The method
is guaranteed to converge to a solution where no further improvement is possible
through single-bit flips. For an example of this type of algorithm applied in a
different context, see \citet{AngerisVB2021}.

In our experiments, we use our method to perform code optimization for two code
design problems. The first is modeled after the GPS L1 C/A codes, which consist
of $63$ length-$1023$ codes. The second is based on the Galileo E1 codes, which
consist of $100$ length-$4092$ codes. We demonstrate that our method converges
to low-correlation codes quickly. Our implementation and experiments have been
made available at \texttt{\url{https://github.com/Stanford-NavLab/decor}}.

The remainder of this paper is structured as follows: \S\ref{s-code-design}
introduces the spreading code design problem. In
\S\ref{s-efficient-computation}, we present efficient methods for evaluating the
impact of altering a single code entry on the correlation matrices and objective
function. \S\ref{s-bit-flip-descent} details the proposed bit-flip descent
method. \S\ref{s-experiments} provides numerical examples, and finally, we
conclude in \S\ref{s-conclusions}.

\section{Spreading code design}\label{s-code-design}

A family of $n$ spreading codes, each of length $T$, is represented by a binary
matrix $X\in\{-1, 1\}^{n\times T}$. Here, $X_{i,t}$ is the $(i,t)$-th bit, or
$t$-th element of the $i$-th spreading code, for $i=1,\ldots,n$ and
$t=0,\ldots,T-1$. The $i$-th spreading code, denoted by $X_i$, corresponds to
the $i$-th row of the matrix $X$.

\subsection{Correlation matrices}

The auto- and cross-correlation properties of the spreading code family $X$ are
described by $T$ correlation matrices, $\Sigma_0, \ldots, \Sigma_{T-1}$, where
each $\Sigma_t \in\reals^{n \times n}$ represents the shift-$t$ correlation
matrix:
\BEQ\label{e-correlation-def}
(\Sigma_t)_{i,j} =
\frac{1}{T}\sum_{\tau=0}^{T-1} X_{i,\tau} X_{j,(\tau-t)_{\bmod T}},
\quad t = 0,\ldots,T-1, \quad i,j = 1,\ldots,n.
\EEQ
The diagonal entries $(\Sigma_t)_{i,i}$ represent the shift-$t$ autocorrelation
of the $i$-th code, while the off-diagonal entries $(\Sigma_t)_{i,j}$ capture
the shift-$t$ cross-correlation between the $i$-th and $j$-th codes, for $i \neq
j$. Note that for any $X$, the shift-zero autocorrelation values are always 1,
\ie, the diagonal entries satisfy $(\Sigma_0)_{i,i} = 1$ for all $i$.

\subsection{Spreading code design problem}

The goal of spreading code design is to find a code family $X$ with small auto-
and cross-correlation values. This may be formulated as an optimization problem
in which we minimize the objective function
\BEQ\label{e-objective}
f(X) = \sum_{(t, i, j) \in \mathcal I}\left|(\Sigma_t)_{i,j}\right|^p,
\EEQ
where $p\ge 1$ is a parameter and 
\[
\mathcal I \subseteq \{0,\ldots,T-1\}\times\{1,\ldots,n\}\times\{1,\ldots,n\}
\]
is an index set. Increasing $p$ increases the penalty on large correlation
values, which can reduce outliers with large magnitude, at the cost of a higher
variance in correlation values.

\PAR{Index set}

Since the cross-correlation is symmetric, we restrict the index set $\mathcal I$
to the indices $j\ge i$. We also exclude the zero-shift autocorrelation values
$(\Sigma_0)_{i,i}$ for $i=1,\ldots,n$, since they do not depend on the value of
$X$. Therefore, we take $\mathcal I$ to be
\[
\mathcal I = \{(t,i,j) \mid j > i\} \cup \{(t,i,i)\mid t > 0\},
\]
which contains $T(n^2+n)/2-n$ indices.

\PAR{Computing the objective}

The objective function \eqref{e-objective} may be directly computed in $O(n^2
T\log T)$ FLOPS (floating-point operations) using the fast Fourier transform
(FFT). Let $\mathcal F_T$ and $\mathcal F_T^{-1}$ denote the forward and inverse
FFT operators, respectively. Then, we have
\BEQ\label{e-correlation-fft}
(\Sigma_t)_{i,j} = 
\mathcal F_T^{-1}(\mathcal F_T(X_i)\circ\mathcal F_T(X_j)^*)_t,
\quad t=0,\ldots,T-1, \quad i,j=1,\ldots,n,
\EEQ
where $\circ$ denotes the element-wise product and $*$ denotes the complex
conjugate. Since $\mathcal F_T$ and $\mathcal F_T^{-1}$ are computable in
$O(T\log T)$ FLOPS using the fast Fourier transform (FFT),
$\Sigma_0,\ldots,\Sigma_{T-1}$ may be computed in $O(n^2 T\log T)$ FLOPS. It
then follows that the objective \eqref{e-cdp} may also be computed in $O(n^2
T\log T)$ FLOPS.

\PAR{Spreading code design problem}

The spreading code design problem is given by the optimization problem
\BEQ\label{e-cdp}
\begin{array}{ll}
\mbox{minimize} & f(X) \\
\mbox{subject to} & X_{i,t}\in\{-1, 1\}, 
\quad i=1,\ldots,n, \quad t=0,\ldots,T-1.
\end{array}
\EEQ
The code design problem \eqref{e-cdp} is a combinatorial optimization problem
that is difficult to solve in general. In \S\ref{s-bit-flip-descent}, we propose
an efficient bit-flip descent method that can quickly generate good solutions to
\eqref{e-cdp}. First, we present methods for efficiently updating the values of
the correlation matrices and objective function, when a single code entry is
changed.

In the following section, we show how the correlation matrices may be updated in
just $O(nT)$ FLOPS after flipping a single bit, without needing to fully
recompute the correlation matrices using \eqref{e-correlation-fft}.

\section{Efficient computation}\label{s-efficient-computation}

In this section, we describe efficient methods for evaluating the effect of
flipping the sign of any bit $X_{a,b}$ on the correlation matrices $(\Sigma_t)$
and the objective function $f(X)$, for $a=1,\ldots,n$ and $b=0,\ldots,T-1$.
These methods form the foundation of the bit-flip descent algorithm introduced
in \S\ref{s-bit-flip-descent}.

\subsection{Updating the correlation matrices}

Suppose we have a code matrix $X$ with correlation matrices $(\Sigma_t)$, which
have already been computed, \eg, using \eqref{e-correlation-fft}. After flipping
the bit $X_{a, b}$, the updated code matrix $X'$ is 
\BEQ\label{e-code-update}
X'_{i,t} = \begin{cases} -X_{i,t} & i=a, t=b, \\
X_{i,t} & \mbox{else}.
\end{cases}
\EEQ
Let $(\Sigma'_t)$ be the correlation matrices of the updated code $X'$. Although
$(\Sigma'_t)$ may be computed in $O(n^2 T\log T)$ FLOPS using
\eqref{e-correlation-fft}, we show how it may be computed in just $O(nT)$ FLOPS
by directly updating $(\Sigma_t)$.

When $X_{a,b}$ is flipped, at most one term in the sum \eqref{e-correlation-def}
changes. Therefore, the updated correlation matrices $(\Sigma'_t)$ may be
computed from $(\Sigma_t)$ using the update rule $\Sigma'_t=g(\Sigma_t, a, b)$,
where
\BEQ\label{e-correlation-update}
g(\Sigma_t, a, b)_{i,j} = 
\begin{cases}
(\Sigma_t)_{i,j} - (2/T)X_{a,b}X_{j,(b+t)_{\bmod T}} & i = a,\;j < a \\
(\Sigma_t)_{i,j} - (2/T)X_{a,b}X_{j,(b-t)_{\bmod T}} & j = a,\;j > a \\
(\Sigma_t)_{i,j} - (2/T)X_{a,b}(X_{a,(b+t)_{\bmod T}} +
X_{a,(b-t)_{\bmod T}}) & i=a,\;j=a,\;t\ne 0 \\
(\Sigma_t)_{i,j} & \mbox{else}.
\end{cases}
\EEQ
The correlation matrices may therefore be computed in $O(nT)$ FLOPS, since $nT$
entries need to be updated, and each update takes $O(1)$ FLOPS. The updates may
also be performed in parallel, since the updates are independent of each other.

\subsection{Computing objective function deltas}

We now show how to efficiently compute the \emph{objective function delta}, or
the change in the objective function value, when $X_{a, b}$ is flipped. Let $X'$
be the updated code matrix \eqref{e-code-update} with $X_{a, b}$ flipped. Then,
the objective function delta corresponding to this bit flip is
\BEQ\label{e-objective-update}
\Delta_{a, b} = f(X') - f(X).
\EEQ
Since the updated correlation matrices $(\Sigma'_t)$ can be computed in $O(nT)$
FLOPS using \eqref{e-correlation-update}, $\Delta_{a,b}$ may also be computed in
$O(nT)$ FLOPS.

Given the current correlation matrices $(\Sigma_t)$, the objective function
delta $\Delta_{a,b}$ is
\BEQ\label{e-delta}
\Delta_{a, b} = 
\sum_{(t,i,j)\in\mathcal J(a)}
\left(\left|g(\Sigma_t, a, b)_{i,j}\right|^p 
- \left|(\Sigma_t)_{i, j}\right|^p \right),
\EEQ
where
\[
\mathcal J(a) = \{(t,i,j) \in \mathcal I \mid i=a\}
\cup \{(t,i,j) \in \mathcal I \mid j=a\}
\]
is the set of indices $(t,i,j)$ for which $i=a$ or $j=a$. This involves a sum
over only $nT-1$ terms, since the terms $(t,i,j)\not\in\mathcal J(a)$ do not
change when $X_{a,b}$ is flipped. Like the correlation matrix updates, the
objective function delta may be computed in parallel, since the terms in the sum
are independent of each other.

\subsection{Updating the objective function delta matrix}

Given a code matrix $X$, the delta matrix $\Delta\in\reals^{n\times T}$ may be
computed for all $nT$ possible bit flips in $O(n^2 T^2)$ FLOPS using
\eqref{e-delta}. In this subsection, we show how the delta matrix may be updated
in only $O(nT^2)$ FLOPS after a single bit flip.

Let $X'$ be the updated code matrix \eqref{e-code-update} after flipping
$X_{a,b}$. Then $X'$ has correlation matrices $(\Sigma_t')$, which may be
computed using the update rule \eqref{e-correlation-update}: $\Sigma'_t =
g(\Sigma_t, a, b)$, for $t=0,\ldots T-1$. Let $\Delta'\in\reals^{n\times T}$ be
the delta matrix for $X'$, \ie, $\Delta'_{a',b'}$ is the change in the objective
function value when $X'_{a',b'}$ is flipped. Then, we may write
$\Delta'_{a',b'}$ as
\BEAS
\Delta'_{a',b'} 
&=& \sum_{(t,i,j)\in\mathcal J(a')}
\left(
|g(\Sigma'_t, a', b')_{i,j}|^p 
- |(\Sigma'_t)_{i, j}|^p
\right) \nonumber\\
&=& \Delta_{a',b'} + 
\sum_{(t,i,j)\in\mathcal J(a')}
\left(
|g(\Sigma'_t, a', b')_{i,j}|^p 
- |(\Sigma'_t)_{i, j}|^p
- |g(\Sigma_t, a', b')_{i,j}|^p 
+ |(\Sigma_t)_{i, j}|^p
\right),
\EEAS
for $a'=1,\ldots,n$ and $b'=0,\ldots,T-1$.

When $a'\ne a$, the sum has only $T$ nonzero terms, corresponding to the cases
where $i=a$ and $j=a'$ or $i=a'$ and $j=a$. Therefore, $\Delta'_{a', b'}$ may be
computed in $O(T)$ FLOPS when $a'\ne a$. When $a'=a$, the sum has no nonzero
terms, and is equivalent to computing directly computing $\Delta'_{a',b'}$ using
\eqref{e-delta} in $O(nT)$ FLOPS.

This leads to the following update rules for $\Delta'_{a',b'}$. If $a'=a$,
\BEQ\label{e-deltas-update-same}
\Delta'_{a',b'} =
\sum_{(t,i,j)\in\mathcal J(a')}
\left(
|g(\Sigma'_t, a', b')_{i,j}|^p 
- |(\Sigma'_t)_{i, j}|^p
\right).
\EEQ
If $a'\ne a$,
\BEQ\label{e-deltas-update-different}
\Delta'_{a',b'} = \Delta_{a',b'} +
\sum_{(t,i,j)\in\mathcal K(a, a')}
\left(
|g(\Sigma'_t, a', b')_{i,j}|^p 
- |(\Sigma'_t)_{i, j}|^p
- |g(\Sigma_t, a', b')_{i,j}|^p 
+ |(\Sigma_t)_{i, j}|^p
\right),
\EEQ
where
\[
\mathcal K(a, a') = \{0,\ldots,T-1\}\times\{(a,a'), (a', a)\}.
\]

Since \eqref{e-deltas-update-same} costs $O(nT)$ FLOPS and
\eqref{e-deltas-update-different} costs $O(T)$ FLOPS, the total cost of updating
$\Delta'$ is $O(nT^2)$ FLOPS. This is a factor of $n$ fewer than recomputing
$\Delta'$ from scratch using \eqref{e-delta}. The delta matrix update may be
done in parallel, since the updates are independent of each other.

In \S\ref{s-bit-flip-descent}, we present a fast bit-flip descent method that
leverages the efficient updates for the correlation matrices, objective function
deltas, and delta matrix derived in this section.

\section{Bit-flip descent}\label{s-bit-flip-descent}

\subsection{Algorithm}

In this section, we propose a fast bit-flip descent method that leverages the
efficient methods for updating the correlation matrices, objective function
deltas, and delta matrix described in \S\ref{s-efficient-computation}. This
enables the method to find good solutions to the spreading code design problem
\eqref{e-cdp} in a reasonable amount of time, even when the number of bits $nT$
is large, on the order of a million.

\PAR{Bit-flip descent}

Starting from an initial code matrix, the bit-flip descent method iteratively
searches for bits that, when flipped, reduce the objective function value. In
the $k$-th iteration, the method randomly selects a set of distinct candidate
bit indices $S_k =\{(a_i,b_i)\}_{i=1}^{M_k}$, where $1\le M_k \le nT$ is the
search size in the $k$-th iteration. The method then selects the index $(a,
b)\in S_k$ that offers the greatest improvement in the objective function value,
\ie, the index that has the smallest delta $\Delta_{a,b}$. If $\Delta_{a,b} <
0$, the bit $X_{a,b}$ is flipped. The method is summarized in Algorithm
\ref{alg-bit-flip-descent}.

\PAR{Convergence}

The method is guaranteed to converge to a local optimum, where no single bit
flip can further reduce the objective function value, although it may not
achieve a global optimum. Nonetheless, the method is highly efficient and, in
practice, consistently generates codes with low auto- and cross-correlation
values. It can be run for a fixed number of iterations $K$, until convergence,
or until a stopping criterion is reached, such as when the objective function
fails to improve after a specified number of iterations.
The method can be carried out multiple times from different random initial 
code designs, after which we take the best one as our design.

\PAR{Initialization}

Algorithm \ref{alg-bit-flip-descent} requires an initial code matrix $X$. In
this work, we initialize $X$ at random, where each entry $X_{i,t}$ takes value
$-1$ or $1$ with equal probability. Alternatively, the method may be initialized
with codes that have been designed using a different method, or with codes that
are known to have good correlation properties, such as the Gold codes
\citep{Gold1967} or the Weil codes \citep{Legendre1808,Rushanan2007}.

\begin{algorithm}[t]
\caption{Bit-flip descent for spreading code design}
\label{alg-bit-flip-descent}
\begin{algorithmic}[1]
\State \textbf{Initialize:} Code matrix $X$
\State Compute correlation matrices $(\Sigma_t)$ using \eqref{e-correlation-fft}
\For{$k = 1, \ldots, K$}
\State Choose a set of $M_k$ distinct indices $S_k =\{(a_i,b_i)\}_{i=1}^{M_k}$ 
at random
\State Select best index $(a, b) = \argmin_{(a',b')\in S^k}\Delta_{a',b'}$
\If{$\Delta_{a, b} < 0$} \label{alg-bit-flip-delta-check}
\State $\Sigma_t \gets g(\Sigma_t, a, b)$ for $t=0,\ldots,T-1$
\State $X_{a, b} \gets -X_{a, b}$
\EndIf
\EndFor
\State \textbf{return} $X$
\end{algorithmic}
\end{algorithm}

\subsection{Choice of search size}

The choice of search size  $M_k$ plays a crucial role in the method's
performance. A larger search size $M_k$ can yield greater improvements in the
objective function with fewer iterations, but comes at the cost of higher
per-iteration computational cost. Conversely, a smaller search size reduces the
cost per iteration but may require more iterations to converge, since it is more
likely to spend time evaluating non-promising bit flips.

We begin by presenting an efficient implementation of the greedy search
strategy, which always selects the bit flip with the best delta and corresponds
to the case where $M_k = nT$. Next, we explore fixed search sizes and introduce
an adaptive strategy that gradually increases the search size $M_k$ over time.

\PAR{Greedy search}

When $M_k = nT$, the method is referred to as a \emph{greedy search}, since the
method always selects the bit flip that leads to the greatest improvement in the
objective function. Greedy search tends to converge in a relatively small number
of iterations, but may be computationally intractable. This is because the
entire delta matrix needs to be computed in each iteration. It costs $O(n^2T^2)$
FLOPS to compute the delta matrix using \eqref{e-delta}, since each delta costs
$O(nT)$ FLOPS. However, the per-iteration cost can be reduced by a factor of $n$
to $O(nT^2)$ FLOPS using the delta matrix update rules
\eqref{e-deltas-update-same} -- \eqref{e-deltas-update-different}. Therefore, it
is only necessary to directly compute the delta matrix in $O(n^2T^2)$ FLOPS
once, for the initial code matrix. In subsequent iterations, the $O(nT^2)$
update rules may be applied after each bit flip. Since a bit is always flipped
in each iteration, the per-iteration cost is $O(nT^2)$ FLOPS.

\PAR{Fixed search size}

Since greedy search can be computationally expensive, an alternative is to use a
fixed search size $M_k = M$, where $M < nT$. When $M = 1$, the method accepts
the first bit flip that improves the objective function. For $M > 1$, the method
instead selects the first bit flip that outperforms $M-1$ other randomly chosen
bit flips. In our experiments, a modest search size of $M = 100$ was found to
achieve the best tradeoff between objective improvement and per-iteration cost,
maximizing convergence speed in terms of overall running time.

\PAR{Adaptive search size}

The fixed search size strategy requires selecting a suitable value for $M$,
which may vary depending on $n$ and $T$, and may require extensive tuning to
identify. We propose an adaptive strategy that gradually increases the search
size $M_k$, which can work well for a range of code design problems without
tuning the search size. The adaptive strategy is initialized with $M_1=1$, and
$M_k$ is increased by $1$ when the selected bit flip fails to improve the
objective function for two consecutive iterations, \ie, when $\Delta_{a,b}\ge 0$
in line \ref{alg-bit-flip-delta-check} of Algorithm \ref{alg-bit-flip-descent}
twice in a row. Since the $k$-th iteration costs $O(M_k nT)$ FLOPS, we switch to
the greedy strategy when the per-iteration cost exceeds the greedy strategy's
per-iteration cost, which is $O(nT^2)$ FLOPS. In our experiments, we switch to
the greedy strategy when $M_k = 10T$.

\subsection{Extensions and variations}

In this subsection, we consider a few possible variations on Algorithm
\ref{alg-bit-flip-descent}.

\PAR{Randomization}

Algorithm \ref{alg-bit-flip-descent} only changes the sign of a code entry if
doing so strictly improves the objective. This can lead to the algorithm being
trapped at suboptimal local minima, where the objective value cannot be improved
by changing a single code entry. Randomization may be introduced to escape those
suboptimal local minima. For example, we may run Algorithm
\ref{alg-bit-flip-descent} several times, each time with a different initial
code matrix, and select the best code matrix found. Another approach is
simulated annealing \citep{BertsimasT1993}, which is a randomized version of
Algorithm \ref{alg-bit-flip-descent} that flips $X_{a,b}$ with some positive
probability when $\Delta_{a,b}\ge 0$. Simulated annealing can find solutions
with lower objective function values since it allows for objective-worsening bit
flips to explore the search space, but tends to require careful tuning and be
slow to converge.

\PAR{Constraints}

The method may be extended to handle constraints on the code matrix $X$. For
example, we may require each code to have the same number of $-1$ and $1$
entries \citep{Soualle2005}. Another constraint is the autocorrelation sidelobe
zero property, which is useful for improving signal tracking performance
\citep{WallnerAHR2007}. Codes that satisfy the autocorrelation sidelobe zero
property have shift-1 autocorrelation values $(\Sigma_1)_{i,i}$ that are zero
(or $\pm 1$, when $T$ is odd). A simple approach for incorporating these
constraints is to modify Algorithm \ref{alg-bit-flip-descent} to only consider
bit flips that do not violate the constraints. That is, if flipping $X_{a,b}$
violates the constraints, then $\Delta_{a,b}$ is taken to be infinite
\citep{YangMG2024b}. This approach requires the initial codes to satisfy the
constraints. An appropriate initialization can be found by modifying the
objective function to penalize only the constraint violations, and then running
Algorithm \ref{alg-bit-flip-descent}.

\section{Numerical examples}\label{s-experiments}

\begin{figure}[t]
\centering
\includegraphics[width=0.45\textwidth]{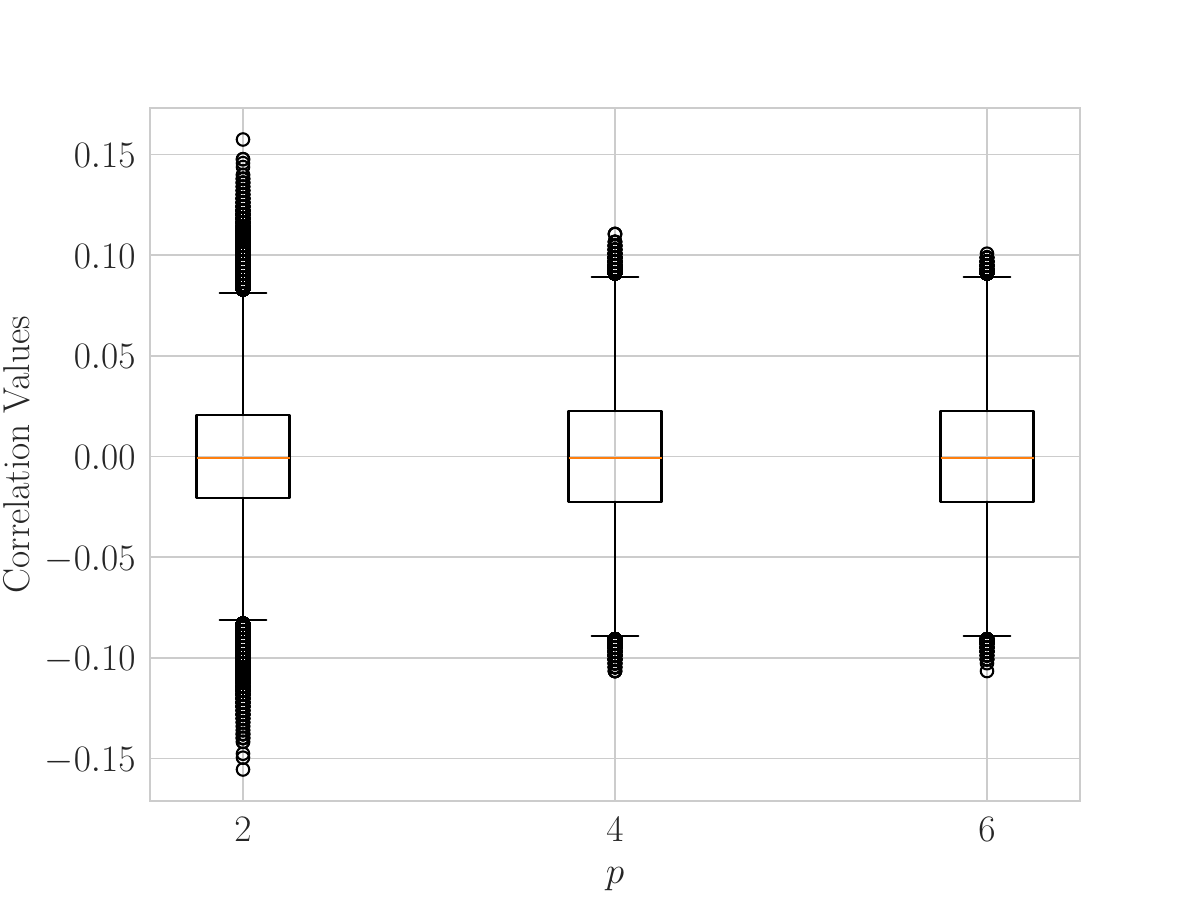}
\caption{Correlation values for varying values of $p$, for the GPS L1 C/A 
test case with $n=63$ and $T=1023$.}
\label{f-p-sweep}
\end{figure}
    
\begin{figure}[t]
\centering
\begin{subfigure}{0.46\textwidth}
\centering
\includegraphics[width=\textwidth]{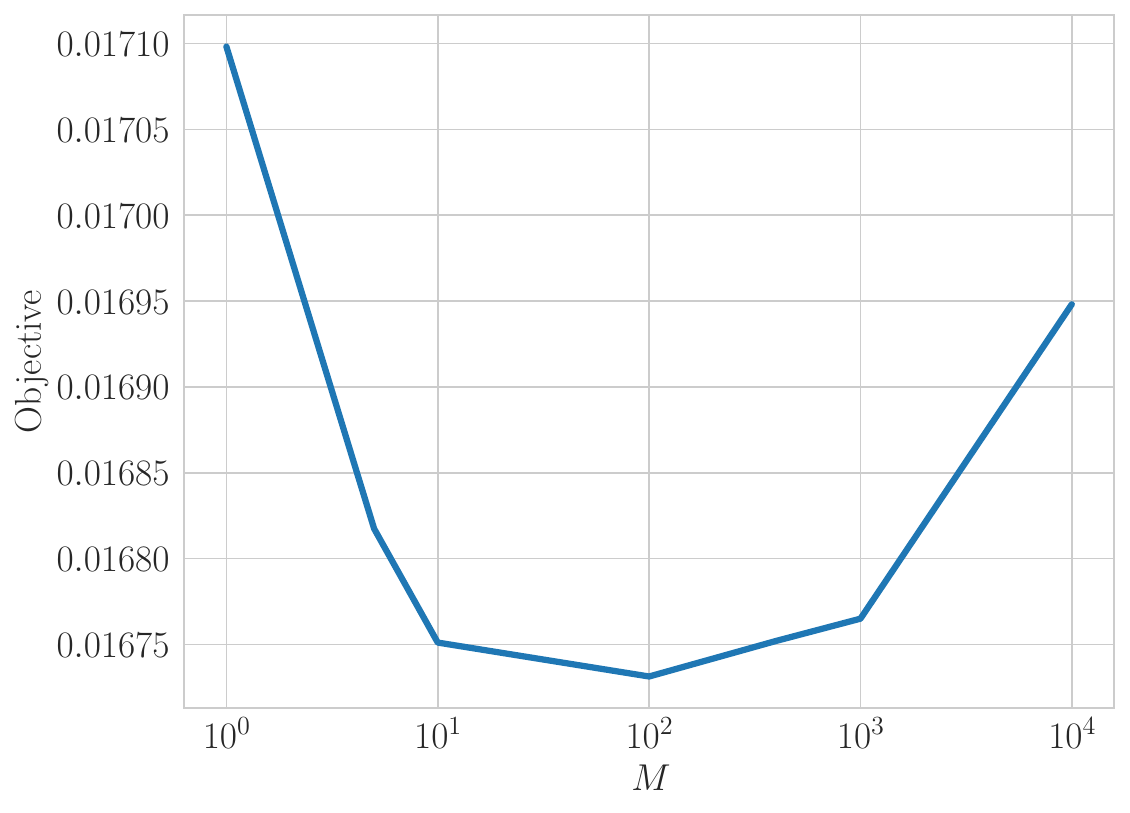}
\caption{GPS L1 C/A: $n=63$, $T=1023$}
\label{f-obj-vs-M-63-1023}
\end{subfigure}
\hfill
\begin{subfigure}{0.46\textwidth}
\centering
\includegraphics[width=\textwidth]{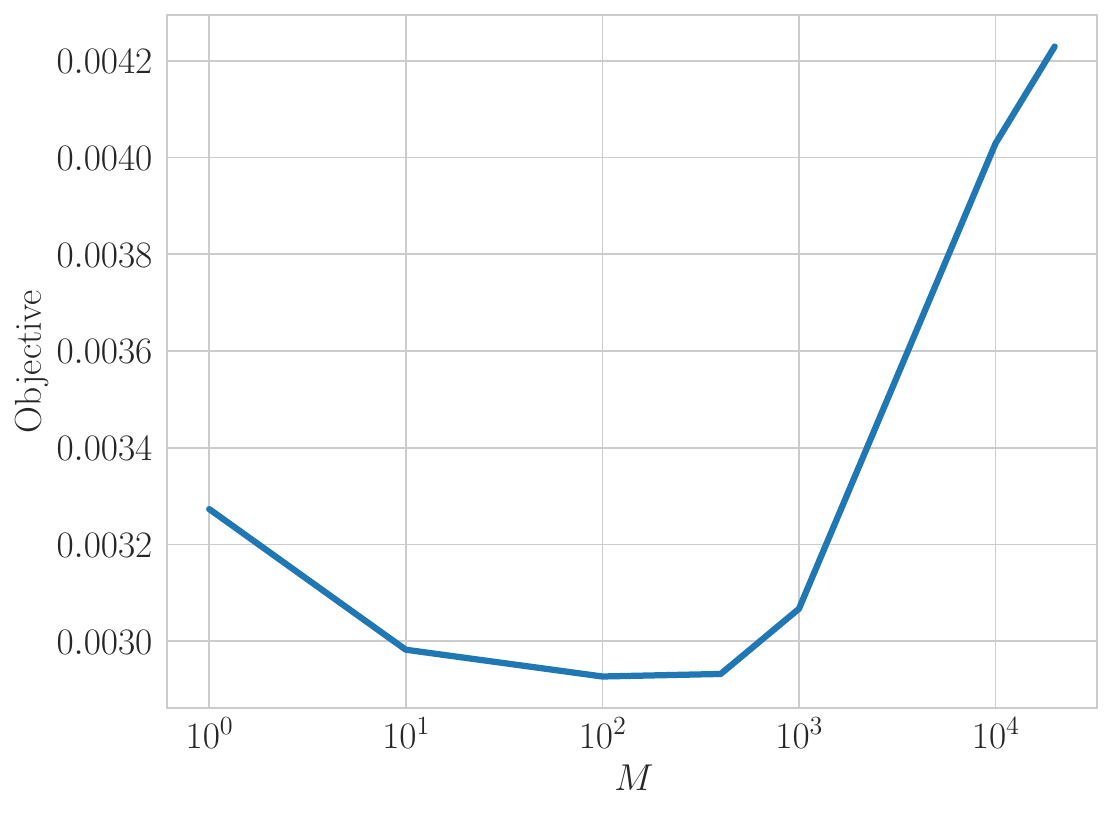}
\caption{Galileo E1: $n=100$, $T=4092$}
\label{f-obj-vs-M-100-4092}
\end{subfigure}
\caption{Objective values achieved in one hour of runtime for different choices of $M$.}
\label{f-obj-vs-M}
\end{figure}

\begin{figure}[t]
\centering
\begin{subfigure}{0.46\textwidth}
\centering
\includegraphics[width=\textwidth]{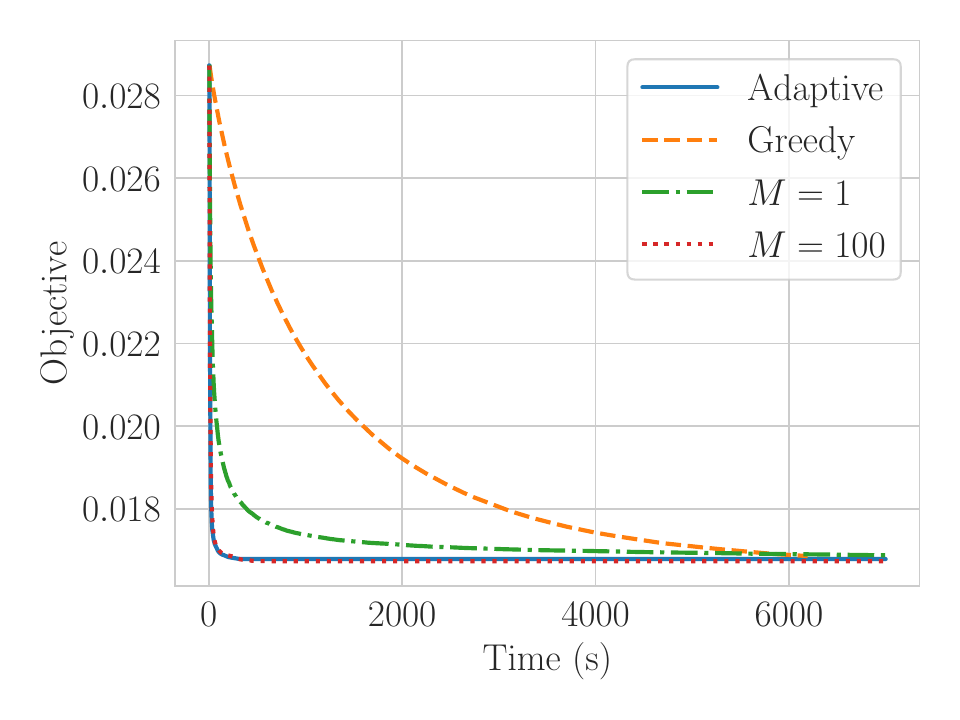}
\caption{GPS L1 C/A: $n=63$, $T=1023$}
\label{f-obj-vs-time-63-1023}
\end{subfigure}
\hfill
\begin{subfigure}{0.46\textwidth}
\centering
\includegraphics[width=\textwidth]{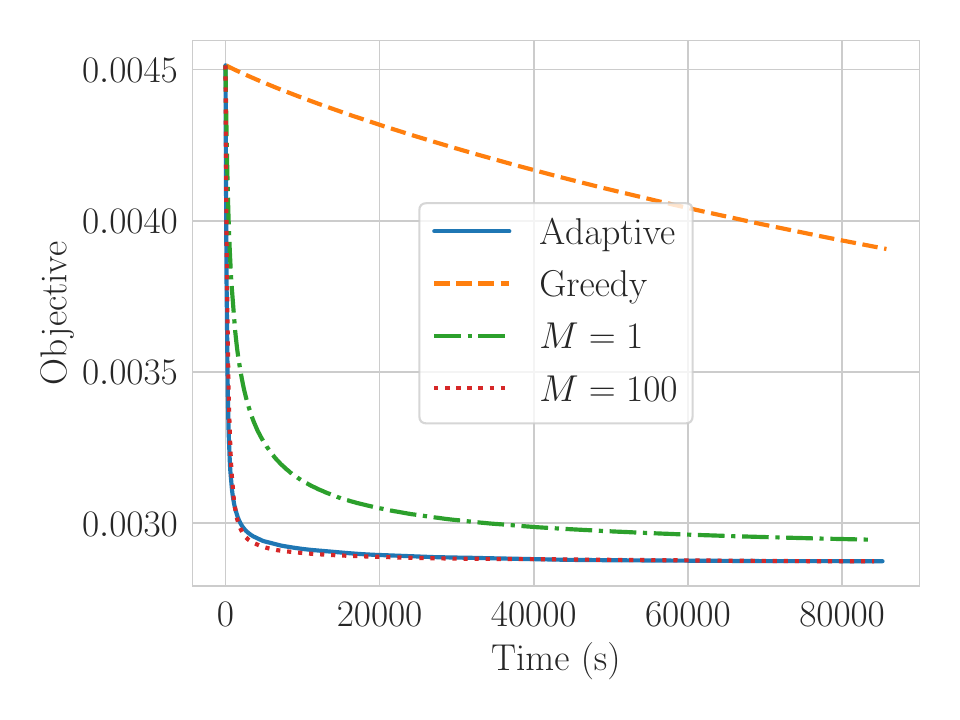}
\caption{Galileo E1: $n=100$, $T=4092$}
\label{f-obj-vs-time-100-4092}
\end{subfigure}\\
\begin{subfigure}{0.46\textwidth}
\centering
\includegraphics[width=\textwidth]{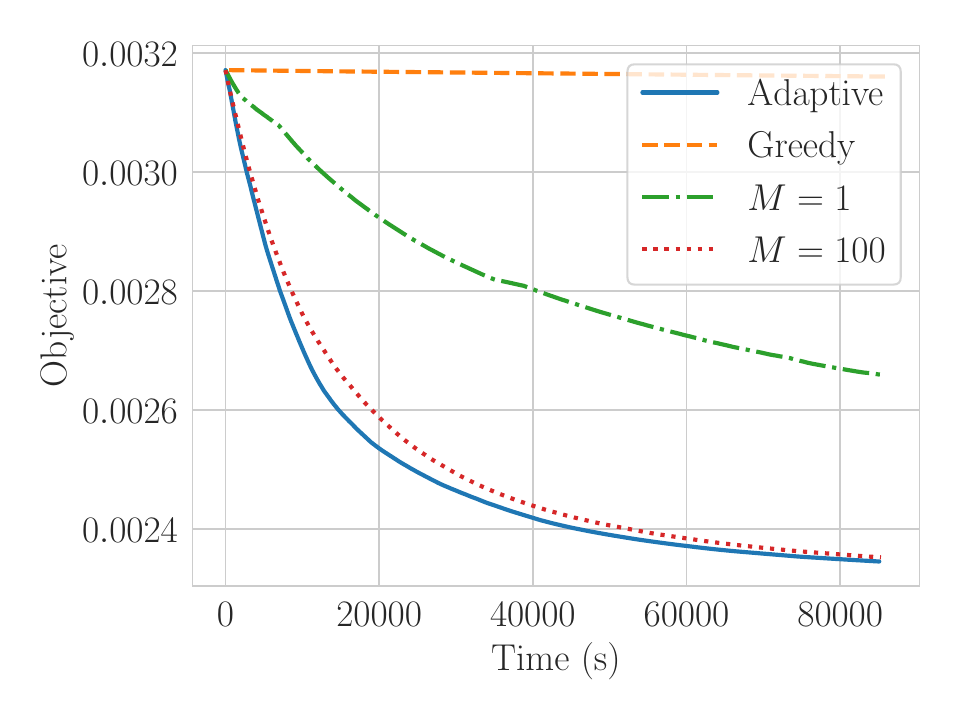}
\caption{GPS L1C: $n=210$, $T=10230$}
\label{f-obj-vs-time-210-10230}
\end{subfigure}
\caption{Objective value vs. runtime of Algorithm \ref{alg-bit-flip-descent}
with four different search strategies, for the three test cases.}
\label{f-obj-vs-time}
\end{figure}

\begin{table}[b]
\centering  
\begin{tabular}{|c|c|c|c|}
\hline
& Code length ($T$) & Family Size ($n$) & Number of bits ($nT$) \\
\hline
GPS L1 C/A & 1,023 & 63 & 64,449 \\
Galileo E1 & 4,092 & 100 & 409,200 \\
GPS L1C & 10,230 & 210 & 2,148,300 \\
\hline
\end{tabular}
\caption{Summary of code design problems.}
\label{t-problems}
\end{table}
    
We consider three code design problems, summarized in Table~\ref{t-problems}.
The first is based on the GPS L1 C/A codes, which consist of $63$ length-$1023$
codes~\citep{interface200gps}. The second is based on the Galileo E1 codes,
which consist of $100$ length-$4092$ codes~\citep{interfaceE1galileo}. The third
and largest code family is based on the GPS L1C signal, and has $210$ codes,
each of length $10230$~\citep{interface800gps}. The total number of bits ranges
from 64,000 for the GPS L1 C/A example to 2 million for the GPS L1C example.

In the following experiments, we initialize the code matrices $X$ at random, and
all methods being compared are initialized with the same code matrix. Unless
otherwise specified, we take $p=6$ in the objective function
\eqref{e-objective}. Each method is run until convergence, or until a time limit
has been reached. The computation of the correlation matrices, objective
function, objective function deltas, and delta matrix were all parallelized. The
GPS L1 C/A and Galileo E1 experiments were run on a machine with a $32$-core
CPU, while the GPS L1C experiments were run on a machine with a $128$-core CPU.
Our code is publicly available at:
\[
\texttt{\url{https://github.com/Stanford-NavLab/decor}}
\]

In \S\ref{s-param-choice}, we evaluate the effect of the parameter $p$ in the
objective function and the search size $M$ in the fixed-size strategy on the
performance of Algorithm \ref{alg-bit-flip-descent}. In \S\ref{s-results}, we
compare the performance of Algorithm \ref{alg-bit-flip-descent} for different
search strategies.

\subsection{Choice of parameters}\label{s-param-choice}

\PAR{Choice of $p$}

In this example, we run Algorithm \ref{alg-bit-flip-descent} with the adaptive
search strategy and three values of $p$, for the GPS L1 C/A test case with
$n=63$ and $T=1023$. Figure \ref{f-p-sweep} shows box plots of the
$T(n^2+n)/2-n$ correlation values $(\Sigma_t)_{i,j}$ of the optimized codes, for
$p=2,4,$ and $6$. Increasing $p$ from $2$ to $6$ reduced the maximum correlation
value from $0.157$ to $0.107$. However, the standard deviation of the
correlation values increased slightly, from $0.0308$ to $0.0310$.

\PAR{Choice of $M$}

Next, we evaluate the effect of the parameter $M$ on the performance of the
fixed search size strategy with $M_k=M$. A good choice of $M$ minimizes the
convergence speed as measured by running time. Increasing $M$ decreases the
number of iterations needed to converge, but also increases the per-iteration
cost. On the other hand, decreasing $M$ reduces the per-iteration cost, but also
increases the number of iterations needed to converge. Figure \ref{f-obj-vs-M}
shows the objective value achieved by Algorithm \ref{alg-bit-flip-descent} for
the GPS L1 C/A and Galileo E1 test cases after one hour of running time, for
different values of $M$. For both test cases, choosing $M=100$ achieves the best
convergence speed.

\subsection{Performance comparison}\label{s-results}

In this subsection, we compare the performance of Algorithm
\ref{alg-bit-flip-descent} for four search strategies: adaptive, greedy search,
and fixed search size with $M=1$ and $M=100$. We ran each method for a maximum
of $24$ hours. For the GPS L1 C/A and Galileo E1 test cases, we ran each method
ten times with different initial codes. For the GPS L1C test case, we ran each
method only once starting from the same random initialization, due to the high
computational cost. Table \ref{t-performance} shows the achieved objective value
and percent improvement over the initial codes for each method and test case.
For the GPS L1 C/A and Galileo E1 test cases, best result out of the ten runs
are reported. Figure \ref{f-obj-vs-time} shows the objective value vs. runtime
for each method and test case, for one of the ten runs. Finally, Table
\ref{t-objective-range} shows the minimum and maximum objective values achieved
by each method over the ten runs.

Overall, the adaptive search strategy achieved the best performance across the
three test cases. The fixed search size strategy with $M=100$ also performed
well, achieving similar performance to the adaptive strategy. While the greedy
strategy performs reasonably for the GPS L1 C/A case, it makes little progress
for the other two test cases, due to the high per-iteration cost. Finally, Table
\ref{t-objective-range} suggests that the particular choice of initial code does
not significantly affect the final objective value, since the minimum and
maximum objective values achieved by each method are similar.

\section{Conclusions}\label{s-conclusions}

In this paper, we introduced a bit-flip descent method for optimizing binary
spreading codes, specifically designed for the large family sizes and extended
lengths required by modern GNSS and emerging PNT applications. We developed
efficient techniques to update correlation matrices, objective functions, and
objective deltas after individual bit flips, allowing the method to scale
effectively to large-scale code design problems involving millions of bits.
Additionally, the method requires minimal tuning, making it a practical and
flexible option for testing various code lengths, family sizes, and objective
functions tailored to specific applications.

\begin{table}[h]
\centering
\begin{tabular}{|l|S[table-format=2.1] S[table-format=2.1]|S[table-format=2.1] S[table-format=2.1]|S[table-format=2.1] S[table-format=2.1]|}
\hline
& \multicolumn{2}{c|}{GPS L1 C/A} & \multicolumn{2}{c|}{Galileo E1} & \multicolumn{2}{c|}{GPS L1C} \\
\cline{2-7}
& \textrm{Obj. ($\times 10^{-3}$)} & \textrm{\% Impr.} & \textrm{Obj. ($\times 10^{-3}$)} & \textrm{\% Impr.} & \textrm{Obj. ($\times 10^{-3}$)} & \textrm{\% Impr.} \\
\hline
Adaptive & 16.68806498168692 & 42.20205524478025 & 2.8671472604796318 & 36.62842493948326 & 2.3458981653237583 & 26.036308986718254 \\
Greedy & 16.760071841447463 & 42.00291183432411 & 3.6604404434443425 & 18.964252328430966 & 3.160830461778593 & 0.3415781809100452 \\
$M=1$ & 16.68241739430621 & 42.12105485025675 & 2.9328426502675153 & 35.13805585422927 & 2.6591199142979143 & 16.160758120013465 \\
$M=100$ & 16.73127237003806 & 42.06710101625448 & 2.8710151558587786 & 36.56490692940683 & 2.352885994225797 & 25.81563796694701 \\
\hline
\end{tabular}
\caption{Objective values and percent improvements over the initial code across the four search strategies and three test cases.}
\label{t-performance}
\end{table}

\begin{table}[h]
\centering
\begin{tabular}{|l|S[table-format=2.1] S[table-format=2.1]|S[table-format=2.1] S[table-format=2.1]|}
\hline
& \multicolumn{2}{c|}{GPS L1 C/A} & \multicolumn{2}{c|}{Galileo E1} \\
\cline{2-5}
& \textrm{Min obj. ($\times 10^{-3}$)} & \textrm{Max obj. ($\times 10^{-3}$)}
& \textrm{Min obj. ($\times 10^{-3}$)} & \textrm{Max obj. ($\times 10^{-3}$)} \\
\hline
Adaptive & 16.68806498168692 & 16.84075793704407 & 2.8671472604796318 & 2.8781030683284405 \\
Greedy & 16.760071841447463 & 16.985622193867684 & 3.6604404434443425 & 3.908508033101301 \\
$M=1$ & 16.68241739430621 & 16.82212241429097 & 2.9328426502675153 & 2.9452576942149866 \\
$M=100$ & 16.73127237003806 & 16.826727906114296 & 2.8710151558587786 & 2.8753167040896117 \\
\hline
\end{tabular}
\caption{Minimum and maximum objective values achieved by each method over ten
runs, for the GPS L1 C/A and Galileo E1 test cases.}
\label{t-objective-range}
\end{table}

\section*{Acknowledgements}
This material is based upon work supported by the Air Force Research Lab (AFRL)
under grant number FA9453-20-1-0002.

% the apacite bibliography style matches the ION bibliography style guidelines.
\bibliographystyle{apalike}
\bibliography{bibliography}

\end{document}